\documentclass[structabstract]{aa}
\usepackage{graphicx}
\usepackage{txfonts}
\usepackage{natbib}

\bibpunct[]{(}{)}{;}{a}{}{,}

\begin{document}
   \title{Spectroscopic study and astronomical detection of\\ doubly 
          $^{13}$C-substituted ethyl cyanide}

   \author{
           L. Margul{\`e}s\inst{1}
           \and
           A. Belloche\inst{2}
           \and
           H.~S.~P. M{\"u}ller\inst{3}
           \and
           R.~A. Motiyenko\inst{1}
           \and
           J.-C. Guillemin\inst{4}
           \and
           R.~T. Garrod\inst{5}
           \and
           K.~M. Menten\inst{2}
           }

   \institute{
              Laboratoire de Physique des Lasers, Atomes, et Mol{\'e}cules, UMR CNRS 8523, 
              Universit{\'e} de Lille I, 59655 Villeneuve d'Ascq C{\'e}dex, France
              \email{laurent.margules@univ-lille1.fr}
              \and
              Max-Planck-Institut f\"ur Radioastronomie, Auf dem H\"ugel 69, 
              53121 Bonn, Germany
              \and
               I.~Physikalisches Institut, Universit{\"a}t zu K{\"o}ln,
              Z{\"u}lpicher Str. 77, 50937 K{\"o}ln, Germany
              \and
              Institut des Sciences Chimiques de Rennes, Ecole Nationale Sup{\'e}rieure de Chimie de Rennes, 
              CNRS, UMR 6226, 11 All{\'e}e de Beaulieu, CS 50837, 35708 Rennes Cedex 7, France
              \and             
              Departments of Chemistry and Astronomy, University of Virginia, Charlottesville, 
              VA 22904, USA
              }

   \date{Received 15 February 2016 / Accepted 9 April 2016}

  \abstract
{We have performed a spectral line survey called EMoCA toward 
Sagittarius~B2(N) between 84.0 and 114.4~GHz with the Atacama Large 
Millimeter/submillimeter Array (ALMA) in its Cycles~0 and 
1. Line intensities of the main isotopic species of ethyl cyanide and its 
singly $^{13}$C-substituted isotopomers observed toward the 
hot molecular core Sagittarius~B2(N2) suggest that the doubly 
$^{13}$C-substituted isotopomers should be detectable also.}
{We want to determine the spectroscopic parameters of all three doubly 
$^{13}$C-substituted isotopologues of ethyl cyanide to search for them in our 
ALMA data.}
{We investigated the laboratory rotational spectra of the three species 
between 150~GHz and 990~GHz. We searched for emission lines produced by these
species in the ALMA spectrum of Sagittarius~B2(N2). We modeled their emission 
as well as the emission of the $^{12}$C and singly $^{13}$C-substituted 
isotopologues assuming local thermodynamic equilibrium.}
{We identified more than 5000 rotational transitions, 
pertaining to more than 3500 different transition frequencies,
in the laboratory for each 
of the three isotopomers. The quantum numbers reach $J \approx 115$ and 
$K_a \approx 35$, resulting in accurate spectroscopic parameters and accurate 
rest frequency calculations beyond 1000~GHz for strong to moderately weak 
transitions of either isotopomer. All three species are unambiguously 
detected in our ALMA data. The $^{12}$C/$^{13}$C column density ratio 
of the isotopomers with one $^{13}$C atom to the ones with two $^{13}$C atoms
is about 25.}
{
Ethyl cyanide is the second molecule after methyl cyanide for which 
isotopologues containing two $^{13}$C atoms have been securely detected in the
interstellar medium. The 
model of our ethyl cyanide data suggests that we should be able to detect 
vibrational satellites of the main species up to at least $\varv_{19} = 1$ at 
$\sim$1130~K and up to $\varv_{13} + \varv_{21} = 2$ at $\sim$600~K for the 
isotopologues with one $^{13}$C atom in our present ALMA data. Such satellites 
may be too weak to be identified unambiguously for isotopologues with two 
$^{13}$C atoms.
}

\titlerunning{Doubly $^{13}$C-substituted ethyl cyanide in the laboratory and in Sgr~B2(N2)}

\maketitle
\hyphenation{For-schungs-ge-mein-schaft}
\hyphenation{wave-guide}

%

\section{Introduction}
\label{intro}

Investigations of the warm (100$-$200~K) and dense parts of high mass 
star-forming regions known as ``hot cores'' (or ``hot corinos'' for their 
low mass star analogs) have unveiled a wealth of 
complex molecules. These are saturated or nearly saturated organic molecules 
for the most part, reaching thus far up to 12 atoms \citep{Belloche14}. Most 
of these molecules have been detected toward \object{Sagittarius~B2} (Sgr~B2
for short). 
The Sgr~B2 molecular cloud complex is one of the most prominent star-forming 
regions in our Galaxy. It is located close to the Galactic centre and contains 
two major sites of high-mass star formation, Sgr~B2(M) and (N). 
\object{Sgr~B2(N)} has 
a greater variety of complex organic molecules than Sgr~B2(M). It contains two 
dense, compact hot cores which are separated by about $5''$ in the north-south 
direction \citep{Belloche08,Qin11,Belloche16}. The more prominent one is 
\object{Sgr~B2(N1)}, 
also known as the Large Molecule Heimat, and \object{Sgr~B2(N2)} is to the 
north of it. 

We used the Atacama Large Millimeter/submillimeter Array (ALMA) in its 
Cycles~0 and 1 to perform a spectral line survey of Sagittarius~B2(N) between 
84.0 and 114.4~GHz. This survey is called Exploring Molecular Complexity with 
ALMA (EMoCA). The high angular resolution ($\sim$1.5--1.8$\arcsec$) achieved 
in the survey allows us to separate the emission of the two hot cores and 
reveals that Sgr~B2(N2) has relatively narrow linewidths ($\sim$5~km~s$^{-1}$), 
which reduces the line confusion compared to previous single-dish surveys of 
Sgr~B2(N). Therefore, our analysis has been focused on this source so far. 
One of the first results of EMoCA has been the first detection in space of a 
branched alkyl molecule, \textit{iso}-propyl cyanide, which is found to be 
nearly as abundant as its straight-chain isomer \textit{n}-propyl 
cyanide \citep{Belloche14}. The laboratory spectroscopic investigation on 
\textit{iso}-propyl cyanide \citep{i-PrCN_rot_2011} was an obvious 
prerequisite. 
The decrease in line confusion, however, was important also because this 
molecule was not detected in our previous single-dish survey of Sgr~B2(N) 
\citep[][]{Belloche13,i-PrCN_rot_2011}.

The lighter ethyl cyanide, C$_2$H$_5$CN, also known as propionitrile, 
is considerably more prominent than either \textit{iso}- or
\textit{n}-propyl 
cyanide in our ALMA survey. In fact, lines of the three isotopomers with 
one $^{13}$C atom are so strong that we expected to be able to detect those 
of the three isotopomers with two $^{13}$C atoms. The parent isotopic species 
was detected in the Orion Molecular Cloud and in Sgr~B2 \citep{EtCN_det_1977} 
soon after the even lighter methyl cyanide was detected \citep{MeCN_det_1971}. 
The molecule was also detected in the hot corinos of low-mass 
protostars such as IRAS~16293-2422, NGC\,1333-IRAS\,2A, and NGC\,1333-IRAS\,4A 
\citep{EtCN_etc_IRAS16293_2003,Taquet15}. 
Rotational transitions within the two lowest-lying vibrational states were 
detected in G327.3$-$0.6 \citep{vib-EtCN_det_2000} and Sgr~B2(N) 
\citep{vib-EtCN_SgrB2_2004}. Transitions of even higher excited states were 
detected in the high-mass star forming regions Orion~KL
and Sgr~B2(N) \citep{high-vib_EtCN_2013,Belloche13}. 
Its $^{13}$C isotopomers were detected first in Orion~IRc2 \citep{13C-EtCN_det_2007} 
and soon thereafter in Sgr~B2(N) \citep{13C-VyCN_2008}. \citet{D_EtCN_15N_rot_2009} 
reported the detection of C$_2$H$_5$C$^{15}$N in Orion~IRc2.

The rotational spectrum of the parent isotopic species has been studied extensively 
in its ground vibrational state. The recent work by \citet{EtCN_rot_2009} extended 
the data to 1.6~THz. Information is also available on some low-lying vibrational 
states \citep{high-vib_EtCN_2013}. The ground vibrational states of all singly 
substituted ethyl cyanide species were studied rather extensively, in particular 
the ones with $^{13}$C \citep{13C-EtCN_det_2007,13C-EtCN_rot_2012}, but also 
the ones with D and $^{15}$N \citep{D_EtCN_15N_rot_2009}.

No data are available for C$_2$H$_5$CN isotopomers with two $^{13}$C atoms. 
Therefore, we prepared ethyl cyanide samples highly enriched in $^{13}$C at 
two positions, recorded its rotational spectra up to 1~THz and searched for 
it in our ALMA data toward Sgr~B2(N2). Details about the laboratory 
experiments and the astronomical observations are given in
Sect.~\ref{exptl}. 
Section~\ref{results} presents the results of the laboratory spectroscopy and 
the analysis of the astronomical spectra. These results are discussed in 
Sect.~\ref{discussion} and the conclusions are given in Sect.~\ref{conclusions}.

\section{Experimental details}
\label{exptl}

\subsection{Synthesis}

Into a three-necked flask equipped with a stirring bar, a reflux condenser, and a 
nitrogen inlet, were introduced triethylene glycol (20~mL), potassium cyanide 
(0.7~g, 10.7~mmol), and iodoethane-$^{13}$C$_2$ (1~g, 6.4~mmol). The mixture was heated 
to 110$^{\circ}{\rm C}$ and stirred at this temperature for one hour. After cooling 
to room temperature, the flask was fitted on a vacuum line equipped with two U-tubes. 
The high boiling compounds were condensed in the first trap cooled at $-30^{\circ}{\rm C}$ 
and propanenitrile-2,3-$^{13}$C$_2$ ($^{13}$CH$_3$$^{13}$CH$_2$CN) was selectively 
condensed in the second trap cooled at $-90^{\circ}{\rm C}$. The reaction was performed 
starting from stoichiometric amounts of 1-iodoethane$^{13}$C and K$^{13}$CN to prepare 
propanenitrile-1,2-$^{13}$C$_2$ (CH$_3$$^{13}$CH$_2$$^{13}$CN), and starting from 
2-iodoethane$^{13}$C and K$^{13}$CN to prepare propanenitrile-1,3-$^{13}$C$_2$ 
($^{13}$CH$_3$CH$_2$$^{13}$CN). The nuclear magnetic resonance data (NMR) 
of the three isotopologues is given in Appendix~\ref{s:NMR}.

\subsection{Lille - submillimeter spectra}

The measurements in the frequency range under investigation (150$-$990~GHz) were 
performed using the Lille spectrometer \citep{Alekseev2012}. A quasi-optic dielectric 
hollow waveguide of 3~m length containing investigated gas at the required pressure 
was used as the sample cell in the spectrometer. The measurements were done at typical 
pressures of 10~Pa and at room temperature. The frequency ranges 150$-$330, 400$-$660, 
and 780$-$990~GHz were covered with various active and passive frequency multipliers 
from VDI Inc. and an Agilent synthesizer (12.5$-$18.25~GHz) was used as the source 
of radiation. Estimated uncertainties for measured line frequencies are 30~kHz and 
50~kHz depending on the observed S/N and the frequency range.

\subsection{Observations}

Part of the observations used in this article have been briefly described in 
\citet{Belloche14}. Here we use the full dataset of the EMoCA survey to search
for the emission of the three doubly $^{13}$C-substituted isotopologues of 
ethyl cyanide toward Sgr~B2(N2) at the equatorial position 
($\alpha, \delta$)$_{\rm J2000}$ =
($17^{\rm h}47^{\rm m}19.86^{\rm s}, -28^\circ22'13.4''$). In brief, the 
survey covers the frequency range 84.1 -- 114.4~GHz with a spectral resolution
of 488.3 kHz (1.7 to 1.3 km~s$^{-1}$). The angular resolution ranges from 
1.4$\arcsec$ to 2.1$\arcsec$.
A detailed account of the observations, 
reduction, and analysis method of the full dataset is reported in 
\citet{Belloche16}.

\section{Results}
\label{results}

\subsection{Laboratory spectroscopy}
\label{obs}

Ethyl cyanide is an asymmetric top rotor with $\kappa = (2B - A - C)/(A - C) = -0.9591$ 
for the parent isotopic species, quite close to the prolate limit of $-1$. The cyano 
group causes a large dipole moment of 3.816~(3)~D along the $a$-inertial axis and a 
still sizeable 1.235~(1)~D along the $b$-inertial axis \citep{VyCN_EtCN_dip_2011}. 
As a consequence $a$-type transitions dominate the room temperature rotational spectrum 
up to about 0.75~THz. Internal rotation splitting of the CH$_3$ group or hyperfine 
structure splitting caused by the $^{14}$N nucleus are only resolvable in selected 
transitions at low frequencies. 
Both types of splitting were not resolved here.
Heavy atom substitution changes the spectroscopic parameters only slightly, and the 
changes in the dipole moment components are very small, such that they are usually 
neglected.


\begin{table*}
\begin{center}
\caption{Spectroscopic parameters of three ethyl 
         cyanide isotopomers with two $^{13}$C atoms
         in comparison to those of the main isotopologue.}
\label{spec-parameter}
\renewcommand{\arraystretch}{1.10}
\vspace*{-2.5ex}
\begin{tabular}[t]{lr@{}lr@{}lr@{}lr@{}l}
\hline \hline
Parameter & \multicolumn{2}{c}{CH$_3$CH$_2$CN\tablefootmark{a}} & \multicolumn{2}{c}{CH$_3^{13}$CH$_2^{13}$CN} 
& \multicolumn{2}{c}{$^{13}$CH$_3$CH$_2^{13}$CN} & \multicolumn{2}{c}{$^{13}$CH$_3^{13}$CH$_2$CN} \\
\hline
$A$                      &  27663&.68206~(52)  &  27021&.85940~(20)  &  27314&.27505~(23)  &  26713&.80045~(26)  \\
$B$                      &   4714&.187784~(80) &   4673&.123591~(27) &   4572&.983663~(31) &   4584&.329843~(36) \\
$C$                      &   4235&.085063~(74) &   4186&.641599~(26) &   4112&.858906~(28) &   4107&.933715~(34) \\
$D_K \times 10^3$        &    547&.7770~(29)   &    530&.8139~(12)   &    551&.8269~(18)   &    523&.8704~(14)   \\
$D_{JK} \times 10^3$     &  $-$47&.26453~(43)  &  $-$44&.84446~(15)  &  $-$48&.15240~(17)  &  $-$45&.53332~(18)  \\
$D_J \times 10^3$        &      3&.008009~(37) &      2&.8877592~(81)  &      2&.8936301~(90) &      2&.8561756~(96) \\
$d_1 \times 10^6$        & $-$685&.888~(10)    & $-$674&.1535~(29)     & $-$653&.1202~(41)     & $-$662&.5892~(44)     \\
$d_2 \times 10^6$        &  $-$32&.7755~(35)   &  $-$33&.5854~(11)   &  $-$29&.6589~(20)   &  $-$32&.1002~(24)   \\
$H_K \times 10^6$        &     31&.3192~(97)   &     30&.1526~(36)   &     31&.6084~(71)   &     29&.9891~(37)   \\
$H_{KJ} \times 10^6$     &   $-$1&.58325~(89)  &   $-$1&.52676~(33)  &   $-$1&.57282~(44)  &   $-$1&.58070~(36)  \\
$H_{JK} \times 10^9$     & $-$118&.60~(14)     & $-$109&.219~(44)      & $-$125&.624~(53)      & $-$111&.647~(50)      \\
$H_J \times 10^9$        &      9&.3563~(74)   &      8&.5651~(10)   &      8&.7869~(12)   &      8&.6217~(11)   \\
$h_1 \times 10^9$        &      3&.9036~(30)   &      3&.67178~(50)    &      3&.64423~(70)    &      3&.63986~(68)    \\
$h_2 \times 10^{12}$     &    514&.09~(97)     &    508&.74~(29)     &    451&.63~(49)     &    485&.59~(44)     \\
$h_3 \times 10^{12}$     &     63&.18~(29)     &     65&.67~(10)     &     56&.59~(18)     &     61&.79~(20)     \\
$L_{K} \times 10^9$      &   $-$2&.105~(14)    &   $-$1&.9089~(36)     &   $-$2&.0784~(89)    &   $-$1&.9253~(33)     \\
$L_{KKJ} \times 10^{12}$ &     83&.27~(65)     &     84&.32~(24)     &     81&.30~(34)     &     88&.24~(24)     \\
$L_{JK} \times 10^{12}$  &   $-$6&.57~(16)     &   $-$5&.796~(63)      &   $-$5&.149~(84)     &   $-$4&.572~(64)      \\
$L_{JJK} \times 10^{12}$ &      0&.522~(19)    &      0&.4905~(49)     &      0&.6057~(63)     &      0&.4825~(55)     \\
$L_{J} \times 10^{15}$   &  $-$42&.37~(63)     &  $-$31&.705~(42)      &  $-$33&.246~(54)      &  $-$32&.335~(45)      \\
$l_{1} \times 10^{15}$   &  $-$21&.35~(33)     &  $-$17&.184~(24)      &  $-$17&.442~(34)      &  $-$17&.077~(30)      \\
$l_{2} \times 10^{15}$   &   $-$4&.351~(62)    &   $-$4&.159~(17)    &   $-$3&.696~(28)    &   $-$3&.935~(22)    \\
$l_{3} \times 10^{15}$   &   $-$1&.170~(23)    &   $-$1&.1694~(85)    &   $-$1&.011~(14)    &   $-$1&.074~(14)    \\
$l_{4} \times 10^{15}$   &   $-$0&.1380~(86)   &   $-$0&.1356~(31)   &   $-$0&.1138~(32)   &   $-$0&.1145~(34)   \\
$P_{K} \times 10^{15}$   &    113&.2~(69)      &       &             &       &             &       &             \\
$P_{KKJ} \times 10^{15}$ &       &             &       &             &       &             &       &             \\
$P_{KJ} \times 10^{15}$  &   $-$1&.707~(92)    &   $-$1&.754~(34)    &   $-$1&.834~(48)    &   $-$1&.658~(33)    \\
$P_{JK} \times 10^{15}$  &      0&.1483~(79)   &      0&.0808~(26)   &      0&.0763~(32)   &      0&.0618~(28)   \\
$P_{JJK} \times 10^{18}$ &   $-$5&.12~(90)     &   $-$4&.08~(19)     &   $-$5&.19~(27)     &   $-$3&.17~(22)     \\
$P_{J} \times 10^{18}$   &      0&.207~(20)    &       &             &       &             &       &             \\
$p_{1} \times 10^{18}$   &      0&.083~(11)    &       &             &       &             &       &             \\
 \hline 
$N_{\rm lines}$\tablefootmark{b}  & \multicolumn{2}{c}{---}  &  4586&   & 3686&  & 3580&  \\
$\sigma$\tablefootmark{c} & \multicolumn{2}{c}{---}  &   25&.9           &      24&.8         &        28&.4    \\
$\sigma_{\rm weighted}$\tablefootmark{d} & \multicolumn{2}{c}{---} &                 0&.860          &       0&.824         &       0&.935 \\
\hline
\end{tabular}
\end{center}
\vspace*{-2.5ex}
\tablefoot{
Watson's $S$ reduction has been used in the representation $I^r$. All 
parameters are given in MHz except for the last three lines. Numbers in 
parentheses are one standard deviations in units of the least significant 
figures.
\tablefoottext{a}{\citet{EtCN_rot_2009}.}
\tablefoottext{b}{Number of distinct lines.}
\tablefoottext{c}{Standard deviation of the fit in kHz.}
\tablefoottext{d}{Weighted deviation of the fit.}
}
\end{table*}


The initial predictions were obtained from scaling data calculated ab initio 
to values of known isotopic species. The rotational and quartic 
centrifugal distortion parameters of the doubly $^{13}$C-substituted species 
were calculated with harmonic force field 
calculations at B3LYP/6-311G++(3df,2pd) level. The same type of calculations 
were done for the two mono substituted $^{13}$CH$_3$CH$_2$CN and 
CH$_3$$^{13}$CH$_2$CN species, and the differences between the ab 
initio and experimental values from \citet{13C-EtCN_rot_2012} 
were calculated. The scaling values from $^{13}$CH$_3$CH$_2$CN were added to 
$^{13}$CH$_3$CH$_2$$^{13}$CN and $^{13}$CH$_3$$^{13}$CH$_2$CN parameters, and 
the ones 
of CH$_3$$^{13}$CH$_2$CN to CH$_3$$^{13}$CH$_2$$^{13}$CN. This method permits 
to obtain first predictions better than 10~MHz in the lowest part of our 
frequency range. 
The $^aR$-branch $J=18-17$ pattern was easily recognised in the spectra around 
150~GHz. The assignment procedure was the same for all three species: 
transitions obeying $a$-type selection rules were assigned up to 330~GHz first, 
then $^bR$- and $^bQ$-branch 
transitions from 150 to 330~GHz. After these first steps all quartic and sextic 
distortional parameters could be determined. Transitions from 400 to 990~GHz 
could be assigned subsequently. Transitions with high $K_a$ values were 
difficult 
to assign because these were frequently weaker than transitions pertaining 
to excited vibrational states and were often blended with these. As is usual 
for a molecule of this size, many lines remain unassigned because assignments 
of transitions in excited vibrational states were beyond the scope of the 
present investigation. 
Predictions of the spectra were carried out with SPCAT \citep{spfit_1991}, 
ASFIT \citep{Kisiel2001} was employed for fitting. Even though both $A$ and 
$S$ reduction of the rotational Hamiltonian perform nearly equally well for 
ethyl cyanide \citep{13C-EtCN_rot_2012}, we used the latter (in the $I^r$ 
representation) because the molecule is close to the symmetric prolate limit. 
The final line lists consist of 
$\sim$6800 transitions for the 1,2-substituted isotopomer and $\sim$5500 for 
the 1,3- and 2,3-substituted isotopomers. 
The number of different transition frequencies is smaller, 
$\sim$4600 and $\sim$3600, respectively, because asymmetry 
splitting was frequently not resolved, and in some cases accidentally 
overlapping lines were retained in the fit. The $J$ values reach $\sim$115, 
and $K_a$ values extend to at least 20 for $b$-type transitions and to around 
35 for $a$-type transitions. We determined for each isotopomer a full set of up 
to eighth-order rotational parameters along with three diagonal decic 
parameters. 
They are presented in Table~\ref{spec-parameter} together with values for 
the parent isotopologue from \citet{EtCN_rot_2009}. Predictions of the 
rotational spectra of the three isotopomers will be available in the catalog 
section\footnote{http://www.astro.uni-koeln.de/cdms/entries} of the CDMS; 
the line, parameter, and fit files, along with additional auxiliary files, 
will be provided in the CDMS 
archive\footnote{http://www.astro.uni-koeln.de/site/vorhersagen/catalog/archive/EtCN/}. 
Supplementary text files S1.txt, S2.txt, S3.txt will be available at CDS. 
They contain 
the transitions used in the fit with experimental frequencies, accuracies and 
residuals from the fits. Table~\ref{supplement} in the appendix 
provides guidance on these files.


\begin{table*}
\begin{center}
\caption{Rotational partition function values of three ethyl cyanide 
         isotopomers with two $^{13}$C atoms at selected temperatures in 
         comparison to those of the main isotopologue.}
\label{Q-values}
\vspace*{-2.5ex}
\begin{tabular}[t]{r@{}lr@{}lr@{}lr@{}lr@{}l}
\hline \hline
\multicolumn{2}{c}{$T$ (K)} & \multicolumn{2}{c}{$Q$(CH$_3$CH$_2$CN)\tablefootmark{a}} & \multicolumn{2}{c}{$Q$(CH$_3^{13}$CH$_2^{13}$CN)} 
& \multicolumn{2}{c}{$Q$($^{13}$CH$_3$CH$_2^{13}$CN)} & \multicolumn{2}{c}{$Q$($^{13}$CH$_3^{13}$CH$_2$CN)} \\
\hline
 300&.0   &\ \ \ 37424&.5763 &\ \ \ \ 38253&.0668 &\ \ \ \ 38806&.4684 &\ \ \ \ 39214&.7041 \\
 225&.0   &      24286&.9324 &        24824&.3519 &        25183&.2785 &        25448&.2672 \\
 150&.0   &      13209&.5867 &        13501&.7478 &        13696&.8350 &        13840&.9909 \\
  75&.0   &       4667&.9361 &         4771&.1127 &         4839&.9793 &         4890&.9278 \\
  37&.5   &       1651&.0572 &         1687&.5264 &         1711&.8515 &         1729&.8717 \\
  18&.75  &        584&.6478 &          597&.5483 &          606&.1417 &          612&.5216 \\
   9&.375 &        207&.4255 &          211&.9936 &          215&.0286 &          217&.2911 \\
\hline
\end{tabular}
\end{center}
\vspace*{-2.5ex}
\tablefoot{
\tablefoottext{a}{Derived from \citet{EtCN_rot_2009}.}}
\end{table*}

Rotational partition function values of ethyl cyanide and its 
three isotopomers containing two $^{13}$C atoms
are provided at selected temperatures in Table~\ref{Q-values}. The 
temperatures are the standard temperatures in the CDMS 
\citep{CDMS_1,CDMS_2} and JPL \citep{JPL-catalog_1998} catalogs. At the elevated temperatures 
in hot cores, such as Sgr~B2(N2), and under the assumption of LTE a considerable part of larger organic molecules is excited to vibrational levels higher than the ground 
vibrational state. At 150~K a vibrational state at an energy of 480~cm$^{-1}$ (or 691~K) has 
a population of 0.01 with respect to $\varv = 0$. The fundamental vibrations of ethyl cyanide 
have been well determined experimentally \citep{EtCN_isos_IR_1981}, however, only a selection 
of overtone and combination levels are known. Therefore, we used the
harmonic oscillator 
approximation, as is commonly done, to evaluate contributions of such states. 
Use of the anharmonic fundamentals accounts in part for the anharmonicity of the vibrations.
The resulting vibrational correction factors to the rotational partition function of the main isotopologue  are given in 
Table~\ref{F_vib-values}. Isotopic differences are small and most likely within the errors of 
the harmonic oscillator approximation for such a heavy molecule as ethyl cyanide. 


\begin{table}
\begin{center}
\caption{Vibrational correction factors to the rotational partition 
         function of ethyl cyanide at selected temperatures.}
\label{F_vib-values}
\vspace*{-2.5ex}
\begin{tabular}[t]{r@{}lr@{}l}
\hline \hline
\multicolumn{2}{c}{$T$ (K)} & \multicolumn{2}{c}{$F_{\rm vib}$} \\
\hline
 300&.0   & 3&.3438 \\
 225&.0   & 2&.1318 \\
 150&.0   & 1&.3848 \\
  75&.0   & 1&.0384 \\
  37&.5   & 1&.0007 \\
  18&.75  & 1&.0000 \\
   9&.375 & 1&.0000 \\
\hline
\end{tabular}
\end{center}
\end{table}


\subsection{Detection toward Sgr~B2(N2)}
\label{obs-astro}

The LTE modelling of the ALMA spectrum of Sgr~B2(N2) reveals the presence of 
the three doubly $^{13}$C-substituted isotopologues of ethyl cyanide: about 
8, 7, and 8 lines that do not suffer too much from blending 
with transitions of other species are clearly detected for 
CH$_3$$^{13}$CH$_2$$^{13}$CN (Fig.~\ref{f:spec_c2h5cn_13c213c3_ve0}), 
$^{13}$CH$_3$CH$_2$$^{13}$CN (Fig.~\ref{f:spec_c2h5cn_13c113c3_ve0}), 
and $^{13}$CH$_3$$^{13}$CH$_2$CN (Fig.~\ref{f:spec_c2h5cn_13c113c2_ve0}), 
respectively. The parameters of these detected lines are listed in 
Tables~\ref{t:list13c213c3}, \ref{t:list13c113c3}, and \ref{t:list13c113c2},
respectively. These tables also contain a few additional lines that we have not 
counted as formally detected because their wings suffer a bit more from 
blending with emission from other species (line(s) around 
106434~MHz for CH$_3$$^{13}$CH$_2$$^{13}$CN, 
86938~MHz and 103906~MHz for $^{13}$CH$_3$CH$_2$$^{13}$CN, and
85727~MHz and 109313~MHz for $^{13}$CH$_3$$^{13}$CH$_2$CN). 
They are used in the population diagrams described below.

\begin{table*}
 {\centering
 \caption{
 Selection of lines of CH$_3$$^{13}$CH$_2$$^{13}$CN covered by the EMoCA survey of Sgr B2(N2).
}
 \label{t:list13c213c3}
 \vspace*{0.0ex}
 \begin{tabular}{lrcccrccrrcrr}
 \hline\hline
 \multicolumn{1}{c}{Transition\tablefootmark{a}} & \multicolumn{1}{c}{Frequency} & \multicolumn{1}{c}{Unc.\tablefootmark{b}} & \multicolumn{1}{c}{$E_{\rm up}$\tablefootmark{c}} & \multicolumn{1}{c}{$g_{\rm up}$\tablefootmark{d}} & \multicolumn{1}{c}{$A_{\rm ul}$\tablefootmark{e}} & \multicolumn{1}{c}{$\sigma$\tablefootmark{f}} & \multicolumn{1}{c}{$\tau_{\rm peak}$\tablefootmark{g}} & \multicolumn{2}{c}{Frequency range\tablefootmark{h}} & \multicolumn{1}{c}{$I_{\rm obs}$\tablefootmark{i}} & \multicolumn{1}{c}{$I_{\rm mod}$\tablefootmark{j}} & \multicolumn{1}{c}{$I_{\rm all}$\tablefootmark{k}} \\ 
  & \multicolumn{1}{c}{\scriptsize (MHz)} & \multicolumn{1}{c}{\scriptsize (kHz)} &  \multicolumn{1}{c}{\scriptsize (K)} & & \multicolumn{1}{c}{\scriptsize ($10^{-5}$ s$^{-1}$)} & \multicolumn{1}{c}{\scriptsize (mK)} & & \multicolumn{1}{c}{\scriptsize (MHz)} & \multicolumn{1}{c}{\scriptsize (MHz)} & \multicolumn{1}{c}{\scriptsize (K km s$^{-1}$)} & \multicolumn{2}{c}{\scriptsize (K km s$^{-1}$)} \\ 
 \hline
10$_{2,9}$ -- 9$_{2,8}$ &   88391.979 &   1 &   28 & 21 &  5.8 &  153 &  0.031 &  88391.1 &  88393.5 &  10.7(6)$^\star$ &   6.7 &  11.3 \\ 
10$_{4,7}$ -- 9$_{4,6}$ &   88700.116 &   1 &   41 & 21 &  5.1 &  153 &  0.034 &  88699.3 &  88702.7 &  11.9(7)$^\star$ &  11.0 &  12.2 \\ 
10$_{4,6}$ -- 9$_{4,5}$ &   88701.248 &   1 &   41 & 21 &  5.1 & -- & -- & -- & -- & -- & -- & -- \\ 
10$_{2,8}$ -- 9$_{2,7}$ &   89609.594 &   1 &   28 & 21 &  6.0 &  149 &  0.031 &  89608.6 &  89611.1 &   8.9(5)$^\star$ &   7.2 &  10.3 \\ 
11$_{0,11}$ -- 10$_{0,10}$ &   95857.723 &   1 &   28 & 23 &  7.7 &  100 &  0.038 &  95856.4 &  95859.4 &  12.5(4)$^\star$ &   9.8 &  10.8 \\ 
11$_{5,7}$ -- 10$_{5,6}$ &   97553.481 &   1 &   55 & 23 &  6.5 &  100 &  0.052 &  97552.5 &  97555.0 &  13.6(3)$^\star$ &  12.8 &  13.4 \\ 
11$_{5,6}$ -- 10$_{5,5}$ &   97553.506 &   1 &   55 & 23 &  6.5 & -- & -- & -- & -- & -- & -- & -- \\ 
11$_{1,10}$ -- 10$_{1,9}$ &   99642.907 &   1 &   30 & 23 &  8.6 &  162 &  0.039 &  99641.8 &  99644.7 &   7.5(6)$^\star$ &  10.3 &  11.4 \\ 
12$_{5,8}$ -- 11$_{5,7}$ &  106433.758 &   1 &   60 & 25 &  8.8 &  123 &  0.063 & 106432.3 & 106435.7 &  17.2(4)$^\star$ &  13.6 &  16.0 \\ 
12$_{5,7}$ -- 11$_{5,6}$ &  106433.815 &   1 &   60 & 25 &  8.8 & -- & -- & -- & -- & -- & -- & -- \\ 
12$_{3,9}$ -- 11$_{3,8}$ &  106684.840 &   1 &   43 & 25 &  9.3 &  123 &  0.040 & 106683.9 & 106686.8 &  12.1(4)$^\star$ &   8.5 &   9.0 \\ 
13$_{0,13}$ -- 12$_{0,12}$ &  112704.985 &   2 &   38 & 27 & 12.6 &  242 &  0.049 & 112703.9 & 112706.8 &  18.4(8)$^\star$ &  14.3 &  14.9 \\ 
 \hline
 \end{tabular}
 }\\[1ex] 
 \tablefoot{
 \tablefoottext{a}{Quantum numbers of the upper and lower levels.}
 \tablefoottext{b}{Frequency uncertainty.}
 \tablefoottext{c}{Upper level energy.}
 \tablefoottext{d}{Upper level degeneracy.}
 \tablefoottext{e}{Einstein coefficient for spontaneous emission.}
 \tablefoottext{f}{Measured rms noise level.}
 \tablefoottext{g}{Peak opacity of the synthetic line.}
 \tablefoottext{h}{Frequency range over which the emission was integrated.}
 \tablefoottext{i}{Integrated intensity of the observed spectrum in brightness temperature scale. The statistical standard deviation is given in parentheses in unit of the last digit. Values marked with a star are used in the population diagrams shown in Figs.~\ref{f:popdiag_c2h5cn_13c13c}a and b.}
 \tablefoottext{j}{Integrated intensity of the synthetic spectrum of CH$_3$$^{13}$CH$_2$$^{13}$CN.}
 \tablefoottext{k}{Integrated intensity of the model that contains the contribution of all identified molecules, including CH$_3$$^{13}$CH$_2$$^{13}$CN.}
 }
 \end{table*}

\begin{table*}
 {\centering
 \caption{
 Selection of lines of $^{13}$CH$_3$CH$_2$$^{13}$CN covered by the EMoCA survey of Sgr B2(N2).
}
 \label{t:list13c113c3}
 \vspace*{0.0ex}
 \begin{tabular}{lrcccrccrrcrr}
 \hline\hline
 \multicolumn{1}{c}{Transition\tablefootmark{a}} & \multicolumn{1}{c}{Frequency} & \multicolumn{1}{c}{Unc.\tablefootmark{b}} & \multicolumn{1}{c}{$E_{\rm up}$\tablefootmark{c}} & \multicolumn{1}{c}{$g_{\rm up}$\tablefootmark{d}} & \multicolumn{1}{c}{$A_{\rm ul}$\tablefootmark{e}} & \multicolumn{1}{c}{$\sigma$\tablefootmark{f}} & \multicolumn{1}{c}{$\tau_{\rm peak}$\tablefootmark{g}} & \multicolumn{2}{c}{Frequency range\tablefootmark{h}} & \multicolumn{1}{c}{$I_{\rm obs}$\tablefootmark{i}} & \multicolumn{1}{c}{$I_{\rm mod}$\tablefootmark{j}} & \multicolumn{1}{c}{$I_{\rm all}$\tablefootmark{k}} \\ 
  & \multicolumn{1}{c}{\scriptsize (MHz)} & \multicolumn{1}{c}{\scriptsize (kHz)} &  \multicolumn{1}{c}{\scriptsize (K)} & & \multicolumn{1}{c}{\scriptsize ($10^{-5}$ s$^{-1}$)} & \multicolumn{1}{c}{\scriptsize (mK)} & & \multicolumn{1}{c}{\scriptsize (MHz)} & \multicolumn{1}{c}{\scriptsize (MHz)} & \multicolumn{1}{c}{\scriptsize (K km s$^{-1}$)} & \multicolumn{2}{c}{\scriptsize (K km s$^{-1}$)} \\ 
 \hline
10$_{7,3}$ -- 9$_{7,2}$ &   86929.429 &   1 &   77 & 21 &  2.7 &  142 &  0.051 &  86928.2 &  86931.6 &  13.8(6) &  14.2 &  14.8 \\ 
10$_{7,4}$ -- 9$_{7,3}$ &   86929.429 &   1 &   77 & 21 &  2.7 & -- & -- & -- & -- & -- & -- & -- \\ 
10$_{5,6}$ -- 9$_{5,5}$ &   86930.299 &   1 &   51 & 21 &  4.0 & -- & -- & -- & -- & -- & -- & -- \\ 
10$_{5,5}$ -- 9$_{5,4}$ &   86930.307 &   1 &   51 & 21 &  4.0 & -- & -- & -- & -- & -- & -- & -- \\ 
10$_{8,2}$ -- 9$_{8,1}$ &   86938.144 &   2 &   93 & 21 &  2.0 &  142 &  0.014 &  86937.0 &  86939.5 &   2.9(5)$^\star$ &   3.2 &   3.5 \\ 
10$_{8,3}$ -- 9$_{8,2}$ &   86938.144 &   2 &   93 & 21 &  2.0 & -- & -- & -- & -- & -- & -- & -- \\ 
11$_{0,11}$ -- 10$_{0,10}$ &   94120.285 &   2 &   27 & 23 &  7.2 &  104 &  0.036 &  94119.0 &  94121.9 &   9.2(4)$^\star$ &   6.8 &   7.2 \\ 
11$_{6,6}$ -- 10$_{6,5}$ &   95623.286 &   1 &   67 & 23 &  5.0 &  100 &  0.041 &  95622.0 &  95627.4 &  18.7(5) &  18.8 &  21.8 \\ 
11$_{6,5}$ -- 10$_{6,4}$ &   95623.286 &   1 &   67 & 23 &  5.0 & -- & -- & -- & -- & -- & -- & -- \\ 
11$_{7,4}$ -- 10$_{7,3}$ &   95625.196 &   1 &   82 & 23 &  4.2 & -- & -- & -- & -- & -- & -- & -- \\ 
11$_{7,5}$ -- 10$_{7,4}$ &   95625.196 &   1 &   82 & 23 &  4.2 & -- & -- & -- & -- & -- & -- & -- \\ 
11$_{2,9}$ -- 10$_{2,8}$ &   96704.440 &   2 &   32 & 23 &  7.6 &  155 &  0.036 &  96702.9 &  96706.4 &  11.6(6)$^\star$ &   9.4 &  11.5 \\ 
12$_{2,11}$ -- 11$_{2,10}$ &  103905.981 &   2 &   37 & 25 &  9.5 &  114 &  0.041 & 103905.0 & 103907.9 &  10.9(4)$^\star$ &   8.4 &   8.7 \\ 
12$_{7,5}$ -- 11$_{7,4}$ &  104321.766 &   1 &   87 & 25 &  6.1 &  114 &  0.088 & 104320.6 & 104324.1 &  21.0(4) &  18.9 &  26.3 \\ 
12$_{7,6}$ -- 11$_{7,5}$ &  104321.766 &   1 &   87 & 25 &  6.1 & -- & -- & -- & -- & -- & -- & -- \\ 
12$_{6,7}$ -- 11$_{6,6}$ &  104322.135 &   1 &   72 & 25 &  6.9 & -- & -- & -- & -- & -- & -- & -- \\ 
12$_{6,6}$ -- 11$_{6,5}$ &  104322.135 &   1 &   72 & 25 &  6.9 & -- & -- & -- & -- & -- & -- & -- \\ 
12$_{2,10}$ -- 11$_{2,9}$ &  105682.862 &   2 &   37 & 25 & 10.0 &  123 &  0.041 & 105681.6 & 105685.0 &  10.7(5)$^\star$ &   9.1 &  11.2 \\ 
13$_{2,12}$ -- 12$_{2,11}$ &  112499.968 &   2 &   42 & 27 & 12.2 &  166 &  0.047 & 112498.4 & 112501.8 &  19.7(6)$^\star$ &  13.5 &  16.4 \\ 
 \hline
 \end{tabular}
 }\\[1ex] 
 \tablefoot{
 \tablefoottext{a}{Quantum numbers of the upper and lower levels.}
 \tablefoottext{b}{Frequency uncertainty.}
 \tablefoottext{c}{Upper level energy.}
 \tablefoottext{d}{Upper level degeneracy.}
 \tablefoottext{e}{Einstein coefficient for spontaneous emission.}
 \tablefoottext{f}{Measured rms noise level.}
 \tablefoottext{g}{Peak opacity of the synthetic line.}
 \tablefoottext{h}{Frequency range over which the emission was integrated.}
 \tablefoottext{i}{Integrated intensity of the observed spectrum in brightness temperature scale. The statistical standard deviation is given in parentheses in unit of the last digit. Values marked with a star are used in the population diagrams shown in Figs.~\ref{f:popdiag_c2h5cn_13c13c}c and d.}
 \tablefoottext{j}{Integrated intensity of the synthetic spectrum of $^{13}$CH$_3$CH$_2$$^{13}$CN.}
 \tablefoottext{k}{Integrated intensity of the model that contains the contribution of all identified molecules, including $^{13}$CH$_3$CH$_2$$^{13}$CN.}
 }
 \end{table*}

\begin{table*}
 {\centering
 \caption{
 Selection of lines of $^{13}$CH$_3$$^{13}$CH$_2$CN covered by the EMoCA survey of Sgr B2(N2).
}
 \label{t:list13c113c2}
 \vspace*{0.0ex}
 \begin{tabular}{lrcccrccrrcrr}
 \hline\hline
 \multicolumn{1}{c}{Transition\tablefootmark{a}} & \multicolumn{1}{c}{Frequency} & \multicolumn{1}{c}{Unc.\tablefootmark{b}} & \multicolumn{1}{c}{$E_{\rm up}$\tablefootmark{c}} & \multicolumn{1}{c}{$g_{\rm up}$\tablefootmark{d}} & \multicolumn{1}{c}{$A_{\rm ul}$\tablefootmark{e}} & \multicolumn{1}{c}{$\sigma$\tablefootmark{f}} & \multicolumn{1}{c}{$\tau_{\rm peak}$\tablefootmark{g}} & \multicolumn{2}{c}{Frequency range\tablefootmark{h}} & \multicolumn{1}{c}{$I_{\rm obs}$\tablefootmark{i}} & \multicolumn{1}{c}{$I_{\rm mod}$\tablefootmark{j}} & \multicolumn{1}{c}{$I_{\rm all}$\tablefootmark{k}} \\ 
  & \multicolumn{1}{c}{\scriptsize (MHz)} & \multicolumn{1}{c}{\scriptsize (kHz)} &  \multicolumn{1}{c}{\scriptsize (K)} & & \multicolumn{1}{c}{\scriptsize ($10^{-5}$ s$^{-1}$)} & \multicolumn{1}{c}{\scriptsize (mK)} & & \multicolumn{1}{c}{\scriptsize (MHz)} & \multicolumn{1}{c}{\scriptsize (MHz)} & \multicolumn{1}{c}{\scriptsize (K km s$^{-1}$)} & \multicolumn{2}{c}{\scriptsize (K km s$^{-1}$)} \\ 
 \hline
10$_{0,10}$ -- 9$_{0,9}$ &   85729.089 &   4 &   23 & 21 &  5.5 &  158 &  0.030 &  85727.9 &  85730.9 &  12.2(7)$^\star$ &   7.0 &   8.9 \\ 
10$_{5,6}$ -- 9$_{5,5}$ &   86999.357 &   4 &   50 & 21 &  4.3 &  142 &  0.038 &  86998.1 &  87001.0 &   9.4(6)$^\star$ &   9.2 &   9.3 \\ 
10$_{5,5}$ -- 9$_{5,4}$ &   86999.366 &   4 &   50 & 21 &  4.3 & -- & -- & -- & -- & -- & -- & -- \\ 
10$_{4,7}$ -- 9$_{4,6}$ &   87022.337 &   4 &   40 & 21 &  4.8 &  142 &  0.033 &  87021.5 &  87024.9 &  14.7(6)$^\star$ &  10.8 &  11.6 \\ 
10$_{4,6}$ -- 9$_{4,5}$ &   87023.409 &   4 &   40 & 21 &  4.8 & -- & -- & -- & -- & -- & -- & -- \\ 
10$_{1,9}$ -- 9$_{1,8}$ &   88957.082 &   4 &   25 & 21 &  6.0 &  153 &  0.032 &  88955.7 &  88958.6 &   7.4(6)$^\star$ &   7.1 &   8.9 \\ 
11$_{1,11}$ -- 10$_{1,10}$ &   92623.369 &   4 &   28 & 23 &  6.9 &  117 &  0.035 &  92622.3 &  92625.2 &   8.6(5)$^\star$ &   6.5 &   6.6 \\ 
11$_{3,8}$ -- 10$_{3,7}$ &   95882.623 &   1 &   37 & 23 &  6.6 &  100 &  0.032 &  95881.8 &  95884.3 &  11.9(3) &   7.9 &  10.1 \\ 
11$_{2,9}$ -- 10$_{2,8}$ &   96879.471 &   4 &   32 & 23 &  7.7 &  155 &  0.034 &  96878.3 &  96881.2 &   8.0(6)$^\star$ &   9.1 &   9.3 \\ 
12$_{0,12}$ -- 11$_{0,11}$ &  102350.118 &   4 &   32 & 25 &  9.4 &  141 &  0.042 & 102348.8 & 102352.2 &  13.7(5)$^\star$ &  11.8 &  11.8 \\ 
12$_{2,11}$ -- 11$_{2,10}$ &  103952.061 &   4 &   37 & 25 &  9.6 &  114 &  0.040 & 103950.9 & 103953.8 &  12.0(4)$^\star$ &   8.4 &  10.0 \\ 
13$_{1,13}$ -- 12$_{1,12}$ &  109312.610 &   4 &   38 & 27 & 11.4 &  123 &  0.047 & 109311.1 & 109314.5 &  18.7(4)$^\star$ &  14.2 &  18.5 \\ 
 \hline
 \end{tabular}
 }\\[1ex] 
 \tablefoot{
 \tablefoottext{a}{Quantum numbers of the upper and lower levels.}
 \tablefoottext{b}{Frequency uncertainty.}
 \tablefoottext{c}{Upper level energy.}
 \tablefoottext{d}{Upper level degeneracy.}
 \tablefoottext{e}{Einstein coefficient for spontaneous emission.}
 \tablefoottext{f}{Measured rms noise level.}
 \tablefoottext{g}{Peak opacity of the synthetic line.}
 \tablefoottext{h}{Frequency range over which the emission was integrated.}
 \tablefoottext{i}{Integrated intensity of the observed spectrum in brightness temperature scale. The statistical standard deviation is given in parentheses in unit of the last digit. Values marked with a star are used in the population diagrams shown in Figs.~\ref{f:popdiag_c2h5cn_13c13c}e and f.}
 \tablefoottext{j}{Integrated intensity of the synthetic spectrum of $^{13}$CH$_3$$^{13}$CH$_2$CN.}
 \tablefoottext{k}{Integrated intensity of the model that contains the contribution of all identified molecules, including $^{13}$CH$_3$$^{13}$CH$_2$CN.}
 }
 \end{table*}

\begin{figure}
\centerline{\resizebox{0.95\hsize}{!}{\includegraphics[angle=0]{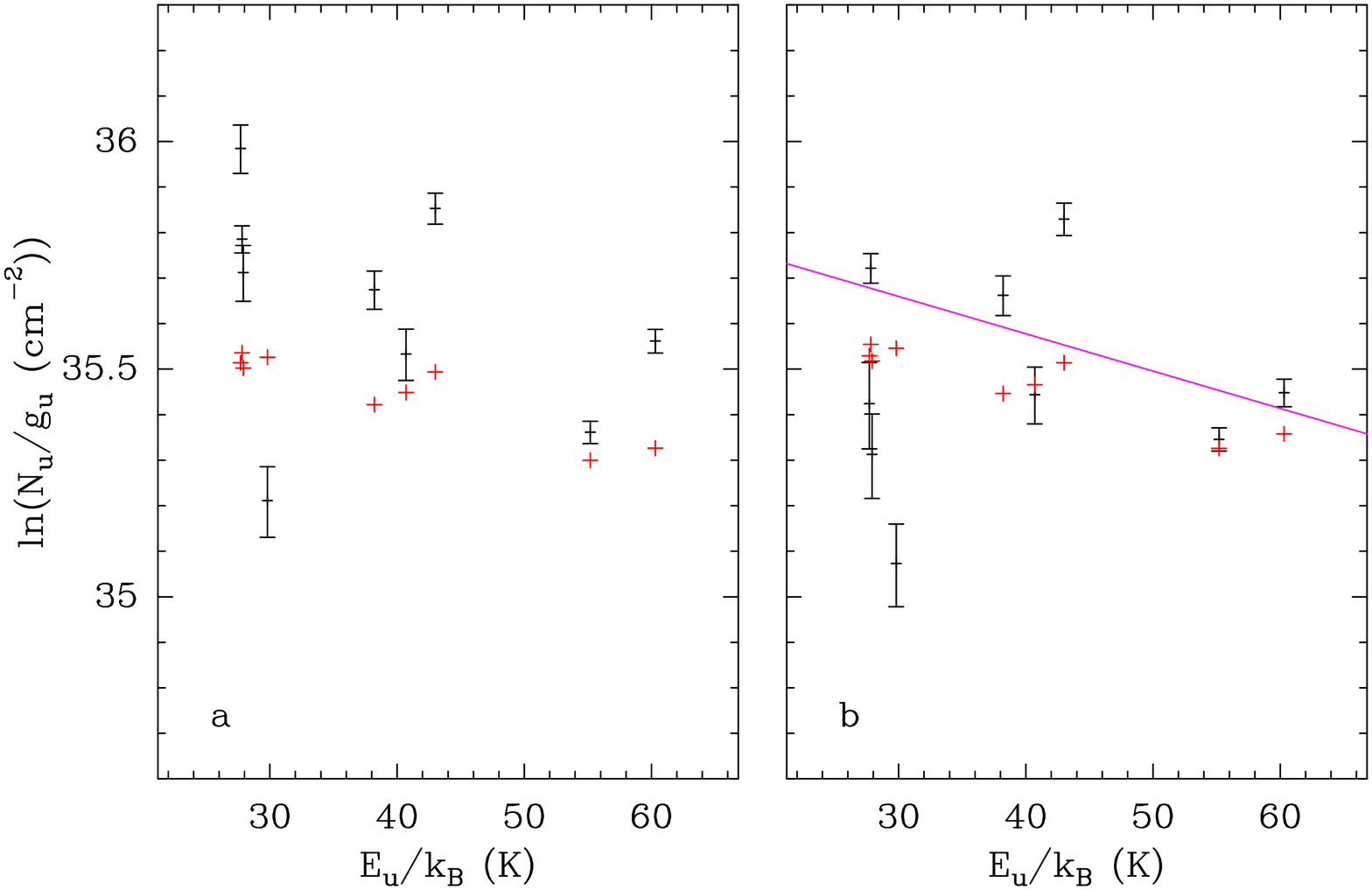}}}
\vspace*{1ex}
\centerline{\resizebox{0.95\hsize}{!}{\includegraphics[angle=0]{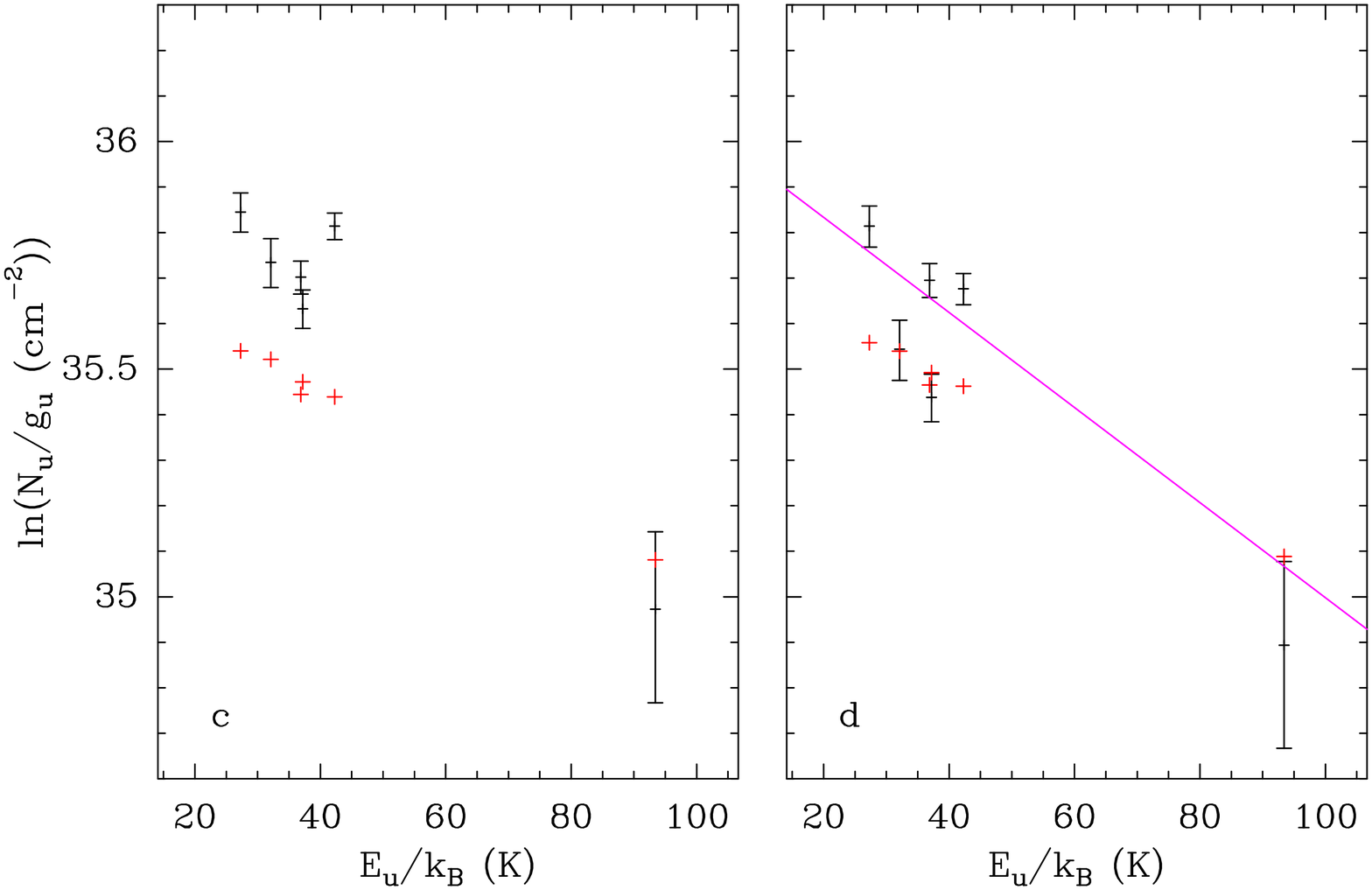}}}
\vspace*{1ex}
\centerline{\resizebox{0.95\hsize}{!}{\includegraphics[angle=0]{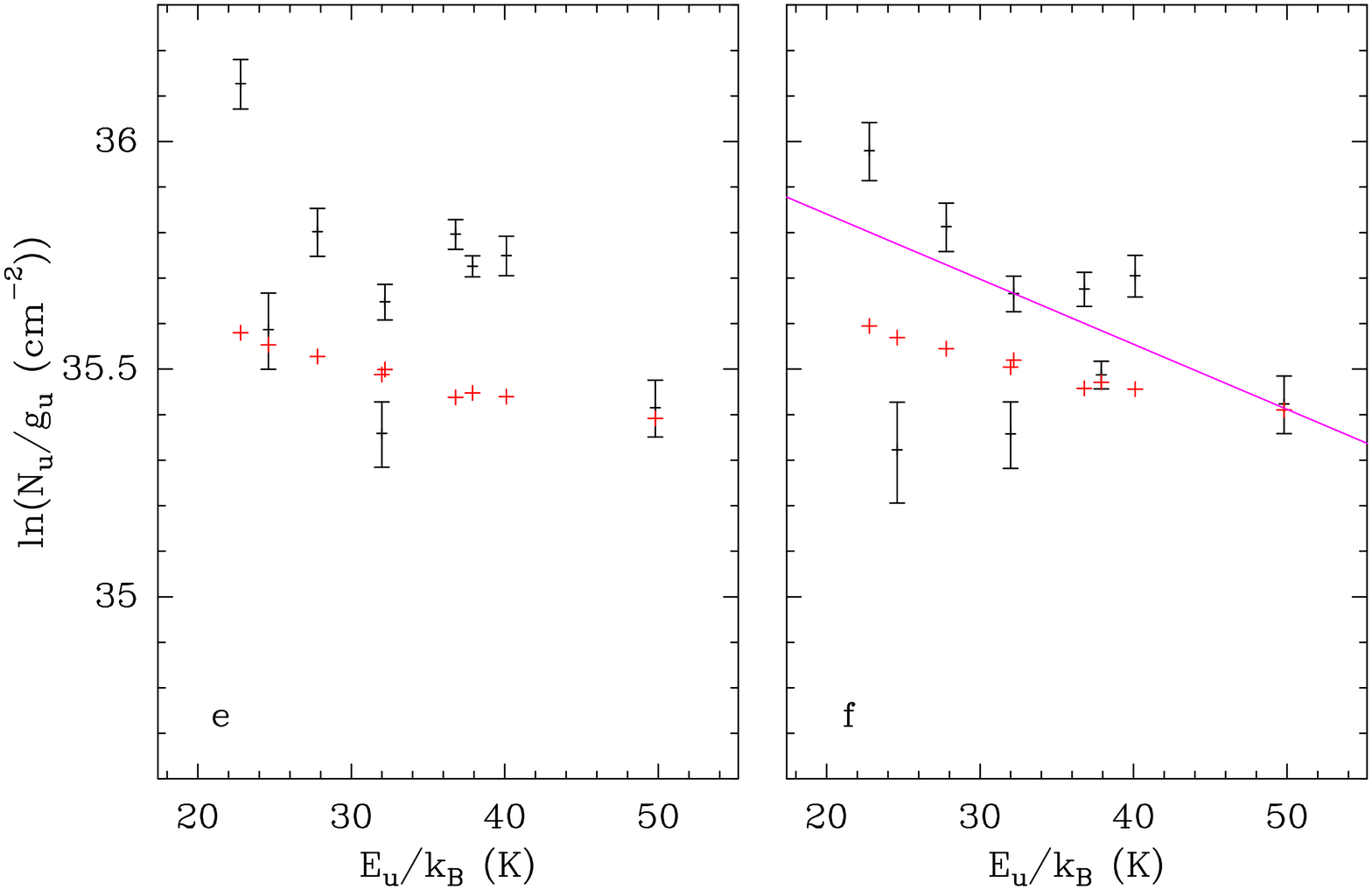}}}
\caption{\textbf{a}, \textbf{b} Population diagram of 
CH$_3$$^{13}$CH$_2$$^{13}$CN, $\varv=0$ toward 
Sgr~B2(N2). Only the lines that are clearly detected and do not suffer too much 
from contamination from other species are displayed. The observed datapoints 
are shown in black while the synthetic populations are shown in red. No 
correction is applied in panel \textbf{a}. In panel \textbf{b}, the optical 
depth correction has been applied to both the observed and synthetic 
populations and the contamination from all other species included in the full 
model has been removed from the observed datapoints. The purple 
line is a linear fit to the observed populations (in linear-logarithmic space).
The derived rotation temperature is given in Table~\ref{t:popfit}.
\textbf{c}, \textbf{d} Same as \textbf{a}, \textbf{b} for 
$^{13}$CH$_3$CH$_2$$^{13}$CN, $\varv=0$.
\textbf{e}, \textbf{f} Same as \textbf{a}, \textbf{b} for 
$^{13}$CH$_3$$^{13}$CH$_2$CN, $\varv=0$.
}
\label{f:popdiag_c2h5cn_13c13c}
\end{figure}

Figure~\ref{f:popdiag_c2h5cn_13c13c} shows the population diagrams of 
the three isotopologues that were constructed using most of the lines listed 
in Tables~\ref{t:list13c213c3}, \ref{t:list13c113c3}, and \ref{t:list13c113c2}.
The number of lines shown in the population diagrams does not exactly 
match the number
of detected lines reported above for several reasons. First of all, 
some detected lines are blends of several transitions of the same molecule 
with different upper-level energies and thus cannot be plotted in a population 
diagram. Second, some lines plotted in the population diagrams are somewhat 
more contamined by other species than what we require to qualify them as 
detected. Still we show them in the population diagrams because we can account 
for this contamination thanks to our full LTE model that contains all species
identified so far.

Table~\ref{t:popfit} lists the results of the rotational temperature fits
for the three doubly $^{13}$C-substituted isotopologues, along with the 
results previously obtained for C$_2$H$_5$CN and the three singly 
$^{13}$C-substituted isotopologues \citep[][]{Belloche16}. The uncertainties 
on the derived rotational temperatures of the doubly $^{13}$C-substituted 
isotopologues are large because of the small number of transitions and the 
narrow range spanned by their upper-level energies. Their rotational 
temperatures are thus poorly constrained but they are consistent within 
1--2$\sigma$ with the value of 150~K that we adopted based on C$_2$H$_5$CN and 
the three singly $^{13}$C-substituted isotopologues \citep[][]{Belloche16}.
Similarly there are too few uncontaminated lines with sufficiently high
signal-to-noise ratios to derive the size of the emission accurately on the
basis of the integrated intensity maps. Therefore we assume the same source
size of 1.2$\arcsec$ as for the more abundant isotopologues. The parameters
of the best-fit LTE models are listed in Table~\ref{t:coldens}. Those for
C$_2$H$_5$CN, the three singly $^{13}$C-substituted isotopologues, and the 
$^{15}$N isotopologue were already reported in \citet{Belloche16}.

Uncertainties are not reported in Table~\ref{t:coldens}. Each 
isotopologue has several lines detected with a signal-to-noise ratio close to
or above 10 if we consider the peak temperatures, or even 20--40 if we consider
the integrated intensities. Therefore, the pure statistical uncertainty on the
derived column densities is smaller than 10\% if we assume that at least one of
the detected lines is free of contamination at its peak or its contamination
is well accounted for by our model (otherwise, the column density can only be 
viewed as an upper limit). \citet{Belloche16} estimated a calibration 
uncertainty of 15\% for the individual setups of the EMoCA survey. Since the 
detected lines are distributed among several setups, we do not expect the 
derived column densities to be biased by the calibration of a particular 
setup. Therefore the column density uncertainty resulting from the calibration 
uncertainty should be less than 15\%. Taking both sources of uncertainty 
together (statistics and calibration), the uncertainty on the column density 
should not be larger than $\sim$15\%. The rotational temperatures were assumed 
to be equal to those of the more abundant isotopologues \citep[][]{Belloche16}. 
Provided this assumption is correct, the uncertainty on the rotational 
temperatures does not have a significant impact on the relative abundance 
ratios which are discussed in Sect.~\ref{astro-discussion}. The same is true 
for the source size and linewidth.

\begin{table}
 {\centering
 \caption{
 Rotational temperatures derived from population diagrams of ethyl cyanide and its isotopologues toward Sgr~B2(N2).
}
 \label{t:popfit}
 \vspace*{0.0ex}
 \begin{tabular}{lll}
 \hline\hline
 \multicolumn{1}{c}{Molecule} & \multicolumn{1}{c}{States\tablefootmark{a}} & \multicolumn{1}{c}{$T_{\rm fit}$\tablefootmark{b}} \\ 
  & & \multicolumn{1}{c}{\scriptsize (K)} \\ 
 \hline
C$_2$H$_5$CN\tablefootmark{c} & $\varv=0$ & 137.3 (1.6) \\ 
$^{13}$CH$_3$CH$_2$CN\tablefootmark{c} & $\varv=0$ & 138.3 (7.5) \\ 
CH$_3$$^{13}$CH$_2$CN\tablefootmark{c} & $\varv=0$ &   112 (11) \\ 
CH$_3$CH$_2$$^{13}$CN\tablefootmark{c} & $\varv=0$ &   150 (40) \\ 
$^{13}$CH$_3$$^{13}$CH$_2$CN & $\varv=0$ &    70 (40) \\ 
$^{13}$CH$_3$CH$_2$$^{13}$CN & $\varv=0$ &    96 (66) \\ 
CH$_3$$^{13}$CH$_2$$^{13}$CN & $\varv=0$ &   122 (77) \\ 
\hline 
 \end{tabular}
 }\\[1ex] 
 \tablefoot{
 \tablefoottext{a}{Vibrational states that were taken into account to fit the population diagram.}
 \tablefoottext{b}{The standard deviation of the fit is given in parentheses. As explained in Sect.~3 of \citet{Belloche16}, these uncertainties should be viewed with caution. They may be underestimated.}
 \tablefoottext{c}{For these species, the analysis was performed in \citet{Belloche16}.}
 }
 \end{table}

\begin{table}[!ht]
 {\centering
 \caption{
 Parameters of our best-fit LTE model of ethyl cyanide and its isotopologues toward Sgr~B2(N2).
}
 \label{t:coldens}
 \vspace*{-1.2ex}
 \begin{tabular}{lrccr}
 \hline\hline
 \multicolumn{1}{c}{Molecule} & \multicolumn{1}{c}{$N_{\rm det}$\tablefootmark{a}} & \multicolumn{1}{c}{$N$\tablefootmark{b}} & \multicolumn{1}{c}{$F_{\rm{vib}}$\tablefootmark{c}} & \multicolumn{1}{c}{$\frac{N_{\rm ref}}{N}$\tablefootmark{d}} \\ 
  & & \multicolumn{1}{c}{\scriptsize (cm$^{-2}$)} & & \\ 
 \hline
 C$_2$H$_5$CN\tablefootmark{e} & 154 &  6.2 (18) & 1.38 &       1 \\ 
 $^{13}$CH$_3$CH$_2$CN\tablefootmark{e} & 54 &  1.9 (17) & 1.38 &      32 \\ 
 CH$_3$$^{13}$CH$_2$CN\tablefootmark{e} & 38 &  1.9 (17) & 1.38 &      32 \\ 
 CH$_3$CH$_2$$^{13}$CN\tablefootmark{e} & 37 &  1.9 (17) & 1.38 &      32 \\ 
 $^{13}$CH$_3$$^{13}$CH$_2$CN & 8 &  7.6 (15) & 1.38 &     818 \\ 
 $^{13}$CH$_3$CH$_2$$^{13}$CN & 8 &  7.6 (15) & 1.38 &     818 \\ 
 CH$_3$$^{13}$CH$_2$$^{13}$CN & 7 &  7.6 (15) & 1.38 &     818 \\ 
 C$_2$H$_5$C$^{15}$N\tablefootmark{e} & 9 &  1.2 (16) & 1.38 &     500 \\ 
\hline 
 \end{tabular}
 }\\[-0.5ex] 
 \tablefoot{
For all species, the model assumes a source size of 1.20$\arcsec$, a rotational temperature of 150~K, a linewidth of 5.0~km~s$^{-1}$, and a velocity offset of -0.8~km~s$^{-1}$ with respect to the  assumed systemic velocity of Sgr~B2(N2), $V_{\mathrm{lsr}} = 74$ km~s$^{-1}$. \tablefoottext{a}{Number of detected lines \citep[conservative estimate, see Sect.~3 of][]{Belloche16}. One line of a given species may mean a group of transitions of that species that are blended together.}
 \tablefoottext{b}{Total column density of the molecule. $X$ ($Y$) means $X \times 10^Y$.}
 \tablefoottext{c}{Correction factor that was applied to the column density to account for the contribution of vibrationally excited states.}
 \tablefoottext{d}{Column density ratio, with $N_{\rm ref}$ the column density of C$_2$H$_5$CN.}
 \tablefoottext{e}{For these species, the analysis was performed in \citet{Belloche16}.}
 }
 \end{table}

\section{Discussion}
\label{discussion}

\subsection{Laboratory spectroscopy}
\label{lab-discussion}

It can be seen in Table~\ref{spec-parameter} that substitution of two of the three $^{12}$C 
atoms in the main isotopologue of ethyl cyanide by $^{13}$C causes only slight decreases 
in the rotational parameters $A$, $B$ and $C$. It is therefore not surprising that the 
magnitudes of the centrifugal distortion parameters also change only slightly, in particular 
those of lower order. It may surprise at first that some of the distortion parameters 
of the doubly $^{13}$C-substituted isotopologues are actually larger than those of the main 
species. One should, however, keep in mind that one finds empirically that the distortion 
parameters scale approximately with appropriate powers of $A - (B + C)/2$, $B + C$ 
and $B - C$. The line lists of the three doubly $^{13}$C-substituted ethyl cyanide 
isotopologues differ somewhat, which easily explains changes in the uncertainties among 
these isotopologues.

\subsection{Astronomical observations}
\label{astro-discussion}

The column densities reported in Table~\ref{t:coldens} imply a 
$^{12}$C$^{12}$C/$^{12}$C$^{13}$C ratio of $\sim$32 and a 
$^{12}$C$^{13}$C/$^{13}$C$^{13}$C ratio of $\sim$25 for ethyl cyanide in 
Sgr~B2(N2).
The $^{12}$C$^{13}$C/$^{13}$C$^{13}$C 
ratio derived for ethyl cyanide is very similar to the 
$^{12}$C$^{13}$C/$^{13}$C$^{13}$C ratio tentatively obtained for HC$_3$N in the 
same source \citep[$\sim$24, see][]{Belloche16} and to the $^{12}$C/$^{13}$C 
ratio derived for methanol and ethanol \citep[$\sim$25, see][]{Mueller16}.
However, the difference on the order of 25\% between the
$^{12}$C$^{12}$C/$^{12}$C$^{13}$C and $^{12}$C$^{13}$C/$^{13}$C$^{13}$C ratios
of ethyl cyanide is a priori surprising. 
Given that the main isotopologue and the singly $^{13}$C-substituted 
ones have a much larger number of detected lines than the doubly 
$^{13}$C-substituted ones, the uncertainty on the column densities of the 
former should be much smaller than the upper limit of 15\% estimated for the 
latter in Sect.~\ref{obs-astro}. Therefore the difference between the two 
ratios seems to be significant. The two ratios would be 
reconciled with a value of 28.6 if the column densities of the singly 
$^{13}$C-substituted isotopologues were higher by a factor 1.12, which would 
correspond to a vertical increment of 0.11 in their population diagrams 
\citep[see Figs.~7b, 8b, and 9b of][]{Belloche16}. Such an increment is not
completely excluded, but it seems to be marginally consistent with the 
observed spectra, given the large number of detected transitions and the fact
that a few of them are already somewhat overestimated by the current model.
The fact that the $^{12}$C$^{12}$C/$^{12}$C$^{13}$C ratio is higher than the
$^{12}$C$^{13}$C/$^{13}$C$^{13}$C ratio cannot be due to an optical depth effect 
either because saturation of ethyl cyanide transitions, if not properly
taken into account, would tend to decrease the apparent 
$^{12}$C$^{12}$C/$^{12}$C$^{13}$C ratio compared to the 
$^{12}$C$^{13}$C/$^{13}$C$^{13}$C ratio. Our analysis takes into account the 
line optical depth and we excluded the lines that are too opaque
\citep[$\tau < 2.5$, see Fig. 6 of][]{Belloche16}, so we believe that the 
$^{12}$C$^{12}$C/$^{12}$C$^{13}$C ratio is not affected by an opacity bias. 
Another possibility is that the column density of the doubly 
$^{13}$C-substituted isotopologues is overestimated by a factor $\sim$1.3.
Given that the EMoCA spectrum is close to the confusion limit, this cannot be
excluded as long as there are still unidentified lines in the survey.

Finally, we cannot exclude that our assumption of a uniform source structure 
and the use of the same source size for all isotopologues introduce systematic
biases in the derived column densities. Figure 10 of \citet{Belloche16} shows 
that the source size actually depends on the upper-level energy of the detected 
transitions of ethyl cyanide. A more elaborated model taking into account the
temperature, density, and possibly abundance gradients in Sgr~B2(N2) would be 
necessary to verify whether the difference between the
$^{12}$C$^{12}$C/$^{12}$C$^{13}$C and $^{12}$C$^{13}$C/$^{13}$C$^{13}$C ratios 
obtained through our simple analysis is significant or not.

\section{Conclusions}
\label{conclusions}

We have unambiguously detected the three isotopomers of ethyl cyanide with two 
$^{13}$C atoms in the Sgr~B2(N2) hot core. Ethyl cyanide is the second 
molecule after methyl cyanide \citep{Belloche16} for which isotopologues 
containing two $^{13}$C atoms have been securely detected in the interstellar
medium. The $^{12}$C/$^{13}$C column density ratio between ethyl 
cyanide isotopomers with one $^{13}$C atom and those with two $^{13}$C atoms is 
$\sim$25, in good agreement with ratios reported for several other 
molecules in this source. 
The $^{12}$C/$^{13}$C ratio between the main isotopologue and the ones with one 
$^{13}$C atom is higher ($\sim$32), but it is unclear at this stage whether
this is a significant difference or a bias due to our simple assumptions about 
the physical structure of the source. The signal-to-noise ratios of the 
detected lines and the derived (rotational) temperature of 150~K suggest that 
vibrational satellites of the isotopologues with two $^{13}$C atoms may be 
just too weak to be identified unambiguously in our current dataset. We 
expect, however, to be able to identify vibrational satellites of the 
isotopologues with one $^{13}$C atom up to the three states with 
$\varv_{13} + \varv_{21} = 2$ at $\sim$600~K, possibly even those of 
$\varv_{12} = 1$ at $\sim$770~K. Vibrational satellites of the main species 
should be observable up to at least $\varv_{19} = 1$ at $\sim$1130~K.


\begin{acknowledgements}
The present investigations were supported by the CNES and the CNRS 
program ``Physique et Chimie du Milieu Interstellaire'' (PCMI).
This work was also done under ANR-13-BS05-0008-02 IMOLABS. Support by 
the Deutsche Forschungsgemeinschaft (DFG) in the framework of the collaborative research 
grant SFB~956, project B3 is also acknowledged. This paper makes use of the following 
ALMA data: ADS/JAO.ALMA\#2011.0.00017.S, ADS/JAO.ALMA\#2012.1.00012.S. ALMA is a 
partnership of ESO (representing its member states), NSF (USA) and NINS (Japan), together 
with NRC (Canada), NSC and ASIAA (Taiwan), and KASI (Republic of Korea), in cooperation 
with the Republic of Chile. The Joint ALMA Observatory is operated by ESO, AUI/NRAO 
and NAOJ. The interferometric data are available in the ALMA archive at 
https://almascience.eso.org/aq/. 
\end{acknowledgements}


\onecolumn
\begin{appendix}
\label{Appendix}

\section{Nuclear magnetic resonance data}
\label{s:NMR}

The NMR type is given first with solvent and the resonance frequency are given in parentheses, 
the shift $\delta$ in ppm with appearance pattern (d, t, q stands for doublet, triplet, 
quartet), the origin of the pattern is given in case of the $^1$H NMR, spin-spin coupling 
parameters $^{n+1}J_{\rm AB}$ in parentheses, the originating molecule group is given at 
the end; A and B are the respective nuclei, and $n$ is the number of atoms between A and B.

\ 

\textbf{Propanenitrile-1,2-$^{13}$C$_2$}

\noindent
$^1$H NMR (CDCl$_3$, 400~MHz): 

\noindent
$\delta$ 1.29 (ddt, 3H, $^2J_{\rm CH} = 4.6$~Hz, $^3J_{\rm CH} = 6.6$~Hz, $^3J_{\rm HH} = 7.7$~Hz, CH$_3$)

\noindent
$\delta$ 2.35 (ddq, 2H, $^1J_{\rm CH} = 134.9$~Hz, $^2J_{\rm CH} = 9.9$~Hz, $^3J_{\rm HH} = 7.7$~Hz, CH$_2$)

\ 

\noindent
$^{13}$C NMR (CDCl$_3$, 100~MHz): 

\noindent
$\delta$ 10.5 (qdd, $^1J_{\rm CH} = 126.1$~Hz, $^1J_{\rm CC} = 33.0$~Hz, $^2J_{\rm CC} = 4.2$~Hz, CH$_3$)

\noindent
$\delta$ 11.1 (td, $^1J_{\rm CH} = 135.0$~Hz, $^1J_{\rm CC} = 56.0$~Hz, CH$_2$)

\noindent
$\delta$ 120.8 (d, $^1J_{\rm CC} = 56.0$~Hz, CN)

\ 

\textbf{Propanenitrile-1,3-$^{13}$C$_2$}

\noindent
$^1$H NMR (CDCl$_3$, 400~MHz): 

\noindent
$\delta$ 1.29 (ddt, 3H, $^1J_{\rm CH} = 126.1$~Hz, $^3J_{\rm CH} = 6.6$~Hz, $^3J_{\rm HH} = 7.7$~Hz, CH$_3$)

\noindent
$\delta$ 2.35 (ddq, 2H, $^2J_{\rm CH} = 5.0$~Hz, $^2J_{\rm CH} = 9.7$~Hz, $^3J_{\rm HH} = 7.7$~Hz, CH$_2$)

\ 

\noindent
$^{13}$C NMR (CDCl$_3$, 100~MHz): 

\noindent
$\delta$ 10.5 (qd, $^1J_{\rm CH} = 126.1$~Hz, $^2J_{\rm CC} = 4.2$~Hz, CH$_3$)

\noindent
$\delta$ 11.1 (tdd, $^1J_{\rm CH} = 135.0$~Hz, $^1J_{\rm CC} = 56.0$~Hz, $^1J_{\rm CC} = 33.0$~Hz, CH$_2$)

\noindent
$\delta$ 120.8 (d, $^2J_{\rm CC} = 4.2$~Hz, CN)

\ 

\textbf{Propanenitrile-2,3-$^{13}$C$_2$}

\noindent
$^1$H NMR (CDCl$_3$, 400~MHz): 

\noindent
$\delta$ 1.30 (ddt, 3H, $^1J_{\rm CH} = 126.1$~Hz, $^2J_{\rm CH} = 4.6$~Hz, $^3J_{\rm HH} = 7.7$~Hz, CH$_3$)

\noindent
$\delta$ 2.35 (dqd, 2H, $^1J_{\rm CH} = 135.0$~Hz, $^3J_{\rm HH} = 7.7$~Hz, $^2J_{\rm CH} = 5.0$~Hz, CH$_2$)

\ 

\noindent
$^{13}$C NMR (CDCl$_3$, 100~MHz): 

\noindent
$\delta$ 10.5 (qd, $^1J_{\rm CH} = 126.1$~Hz, $^1J_{\rm CC} = 33.0$~Hz, CH$_3$)

\noindent
$\delta$ 11.1 (td, $^1J_{\rm CH} = 135.0$~Hz, $^1J_{\rm CC} = 33.0$~Hz, CH$_2$)

\noindent
$\delta$ 120.8 (dd, $^1J_{\rm CC} = 56.0$~Hz, $^2J_{\rm CC} = 4.2$~Hz, CN)

\section{Experimental data}

The experimental transition frequencies of the doubly $^{13}$C-substituted isotopomers
of ethyl cyanide are available as supplementary material. The files S1.txt, S2.txt 
and S3.txt refer to the 1,2-, 1,3- and 2,3-substituted isotopomers, respectively. 
Only the first 10 and the last 11 lines of the 1,2-species appear in
Table~\ref{supplement}; 
for complete versions of all isotopomers see the electronic 
edition\footnote{http://cdsarc.ustrasbg.fr/cgi-bin/VizieR?-source=J/A+A/Vol/Num}. 
The files give the rotational quantum numbers $J$, $K_a$, and $K_c$ 
for the upper state followed by those for the lower state. The observed 
transition frequency is given in megahertz units with its uncertainty and 
the residual between observed frequency and that calculated from the final set 
of spectroscopic parameters. Blended transitions are treated by fitting the 
intensity-averaged frequency, and this weight is also given in the tables. 
In most cases, the blending is caused by unresolved asymmetry splitting, i.e., 
the blended transitions agree in terms of their quantum numbers except for $K_c$ 
(prolate paired transitions) or $K_a$ (oblate paired transitions), and both 
transitions are equal in intensity. Accidental blending of transitions occured 
occasionally.


\begin{table*}
\begin{center}
\caption{Experimental data for the 1,2-isotopomer of ethyl cyanide.}
\label{supplement}
\begin{tabular}{rrrrrrrr@{}lrrr}
\hline \hline
Nr.: & $J'$ & $K_a'$ & $K_c'$ & $J''$ & $K_a''$ & $K_c''$ & 
\multicolumn{2}{c}{Frequency} & Unc. & \multicolumn{1}{c}{O$-$C} & Weight \\
\hline
   1: & 17 &  1 & 17 &  16 &  0 & 16 &  150042&.190 &   0.030 &    0.01469 &        \\
   2: & 33 &  2 & 31 &  33 &  1 & 32 &  150331&.396 &   0.030 &    0.02378 &        \\
   3: & 42 &  5 & 37 &  41 &  6 & 36 &  150403&.376 &   0.030 &    0.04998 &        \\
   4: & 18 &  0 & 18 &  17 &  1 & 17 &  150425&.187 &   0.030 & $-$0.00721 &        \\
   5: & 17 &  8 &  9 &  16 &  8 &  8 &  150780&.154 &   0.030 & $-$0.00667 & 0.5000 \\
   6: & 17 &  8 & 10 &  16 &  8 &  9 &  150780&.154 &   0.030 & $-$0.00667 & 0.5000 \\
   7: & 17 &  9 &  8 &  16 &  9 &  7 &  150784&.377 &   0.030 & $-$0.00577 & 0.5000 \\
   8: & 17 &  9 &  9 &  16 &  9 &  8 &  150784&.377 &   0.030 & $-$0.00577 & 0.5000 \\
   9: & 17 &  7 & 10 &  16 &  7 &  9 &  150789&.125 &   0.030 & $-$0.00886 & 0.5000 \\
  10: & 17 &  7 & 11 &  16 &  7 & 10 &  150789&.125 &   0.030 & $-$0.00886 & 0.5000 \\
      &    &    &    &     &    &    &        &     &         &            &        \\
6810: & 28 & 17 & 11 &  27 & 16 & 12 &  985858&.551 &   0.030 &    0.02324 & 0.5000 \\
6811: & 28 & 17 & 12 &  27 & 16 & 11 &  985858&.551 &   0.030 &    0.02324 & 0.5000 \\
6812: & 64 & 10 & 54 &  63 &  9 & 55 &  986054&.883 &   0.030 & $-$0.00124 &        \\
6813: & 33 & 16 & 17 &  32 & 15 & 18 &  986740&.931 &   0.030 &    0.00118 & 0.5000 \\
6814: & 33 & 16 & 18 &  32 & 15 & 17 &  986740&.931 &   0.030 &    0.00118 & 0.5000 \\
6815: & 48 & 13 & 35 &  47 & 12 & 36 &  987077&.762 &   0.030 & $-$0.01559 & 0.5000 \\
6816: & 48 & 13 & 36 &  47 & 12 & 35 &  987077&.762 &   0.030 & $-$0.01559 & 0.5000 \\
6817: & 38 & 15 & 23 &  37 & 14 & 24 &  987305&.187 &   0.030 & $-$0.07257 & 0.5000 \\
6818: & 38 & 15 & 24 &  37 & 14 & 23 &  987305&.187 &   0.030 & $-$0.07257 & 0.5000 \\
6819: & 43 & 14 & 29 &  42 & 13 & 30 &  987468&.359 &   0.030 &    0.01296 & 0.5000 \\
6820: & 43 & 14 & 30 &  42 & 13 & 29 &  987468&.359 &   0.030 &    0.01296 & 0.5000 \\
\hline
\end{tabular}\\[2pt]
\end{center}
\tablefoot{
The table contains the line number, rotational quantum numbers of the 
assigned transition,
observed transition frequency (MHz), experimental uncertainty (MHz), 
residual O$-$C between observed frequency and that calculated from the final 
set of spectroscopic parameters, and weight for blended lines.
This table as well as those of other conformers are available in their 
  entirety in the electronic edition in the online journal: 
  http://cdsarc.ustrasbg.fr/cgi-bin/VizieR?-source=J/A+A/Vol/Num. 
  A portion is shown here for guidance regarding its form and content.
}
\end{table*}

\section{Additional figures}

\begin{figure*}
\centerline{\resizebox{0.9\hsize}{!}{\includegraphics[angle=0]{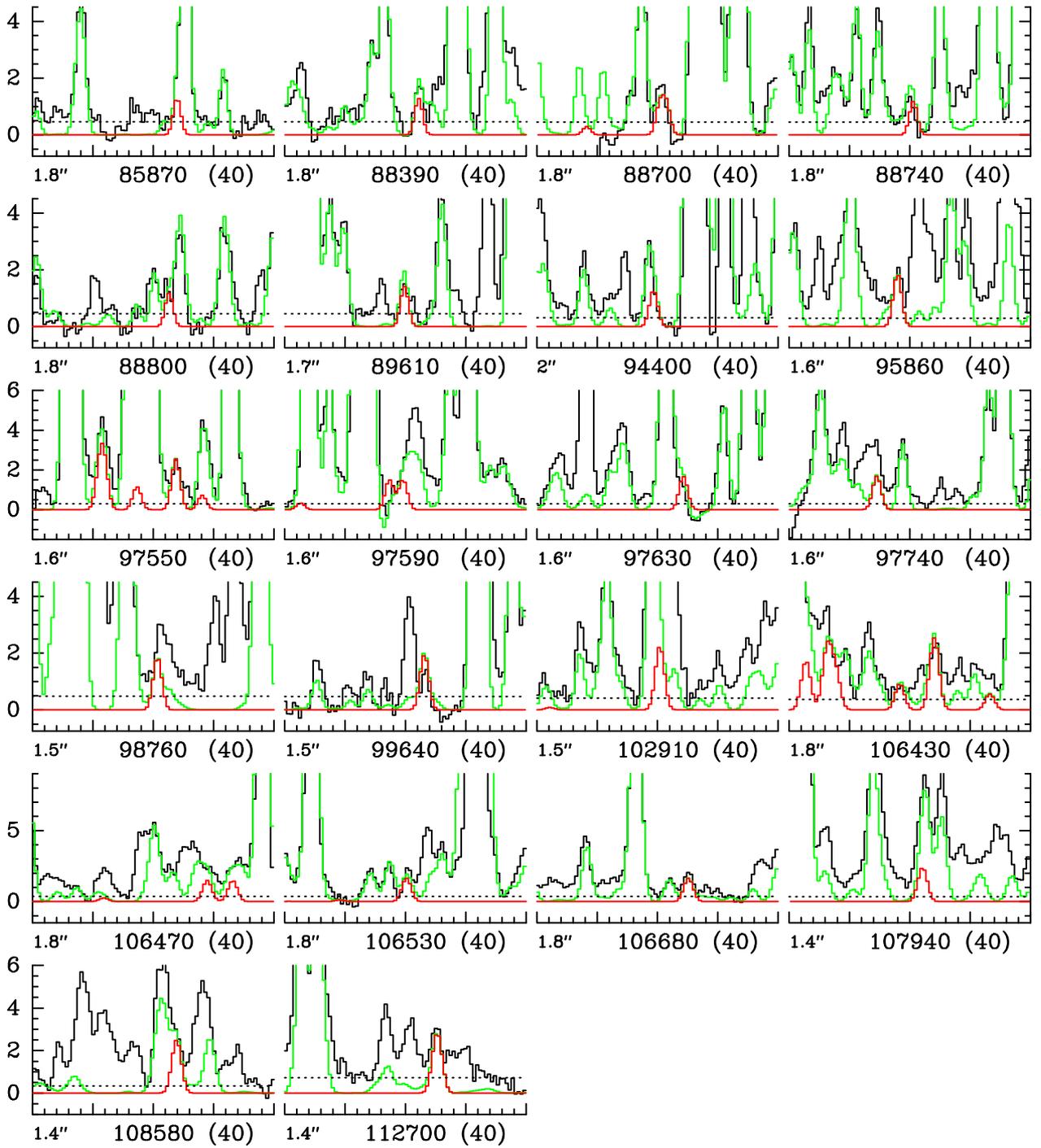}}}
\caption{Transitions of CH$_3$$^{13}$CH$_2$$^{13}$CN, $\varv = 0$ covered by 
our ALMA survey. The best-fit LTE synthetic spectrum of 
CH$_3$$^{13}$CH$_2$$^{13}$CN is displayed in red and overlaid on the observed 
spectrum of Sgr~B2(N2) shown in black. The green synthetic spectrum contains
the contributions of all molecules identified in our survey so far, including 
the one shown in red. The central frequency and width are indicated in MHz 
below each panel. The angular resolution (HPBW) is also indicated. The y-axis 
is labeled in brightness temperature units (K). The dotted line indicates the 
$3\sigma$ noise level.}
\label{f:spec_c2h5cn_13c213c3_ve0}
\end{figure*}

\begin{figure*}
\centerline{\resizebox{0.9\hsize}{!}{\includegraphics[angle=0]{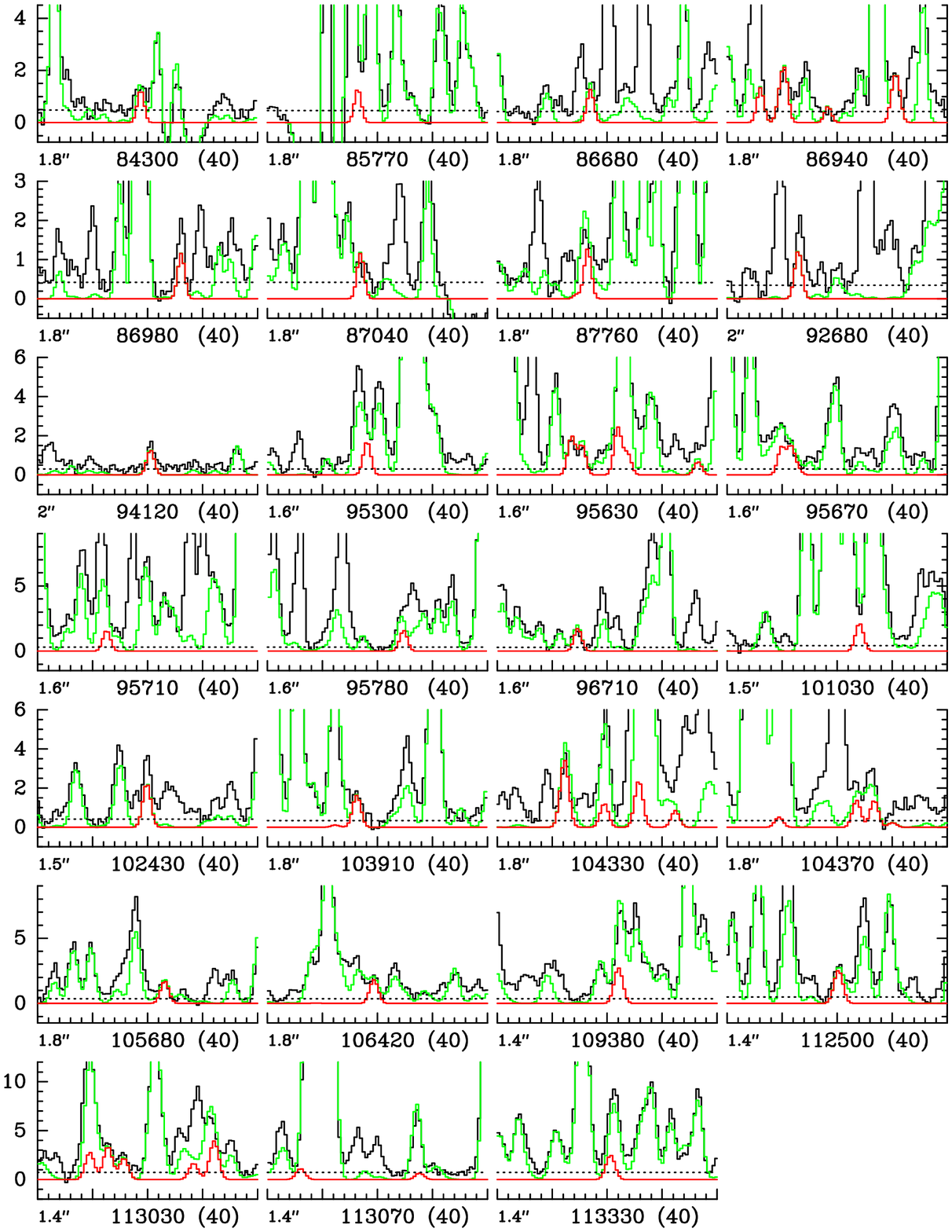}}}
\caption{Same as Fig.~\ref{f:spec_c2h5cn_13c213c3_ve0} for 
$^{13}$CH$_3$CH$_2$$^{13}$CN, $\varv=0$.}
\label{f:spec_c2h5cn_13c113c3_ve0}
\end{figure*}

\begin{figure*}
\centerline{\resizebox{0.9\hsize}{!}{\includegraphics[angle=0]{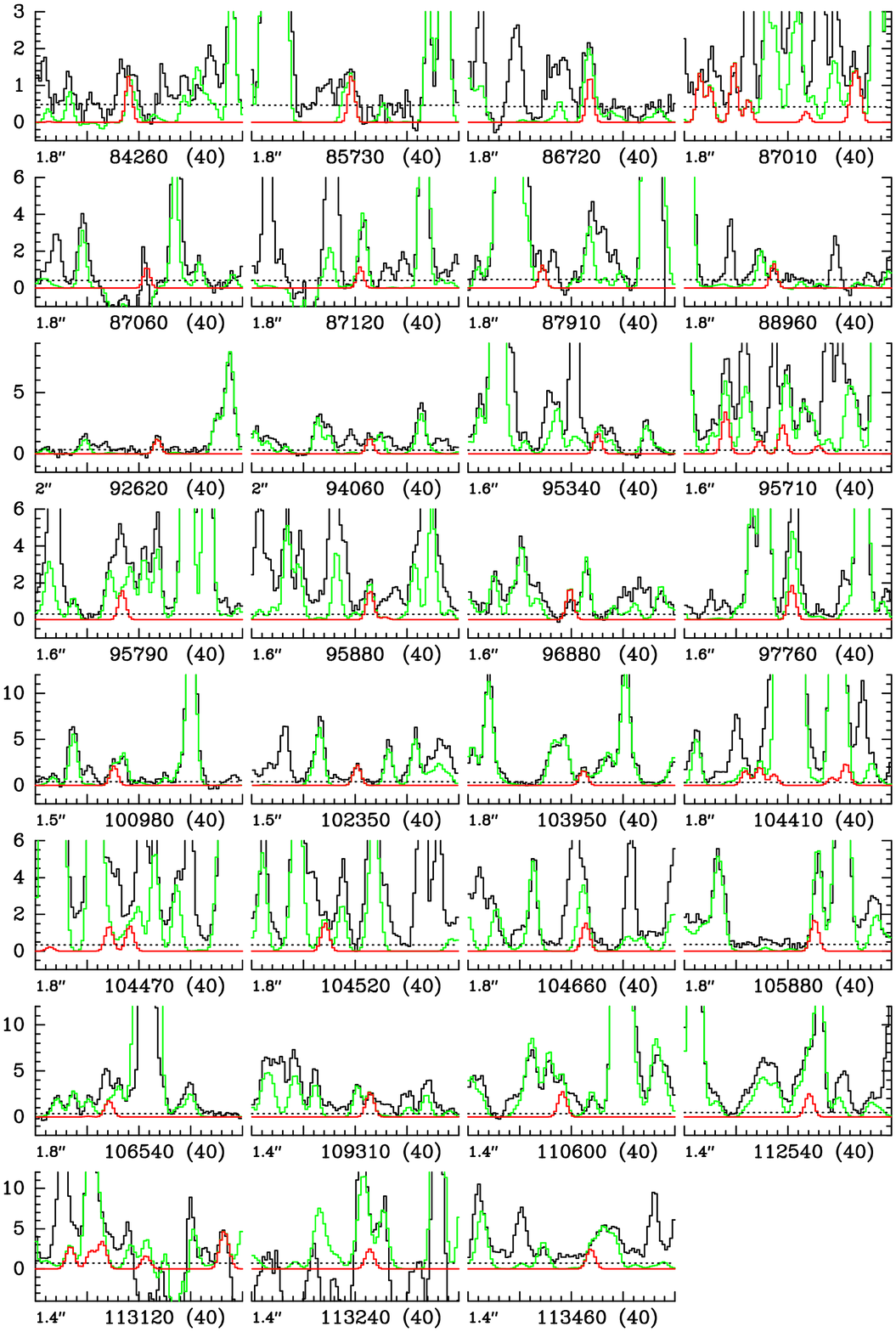}}}
\caption{Same as Fig.~\ref{f:spec_c2h5cn_13c213c3_ve0} for 
$^{13}$CH$_3$$^{13}$CH$_2$CN, $\varv=0$.}
\label{f:spec_c2h5cn_13c113c2_ve0}
\end{figure*}

Figures~\ref{f:spec_c2h5cn_13c213c3_ve0}, \ref{f:spec_c2h5cn_13c113c3_ve0},
and \ref{f:spec_c2h5cn_13c113c2_ve0} show the astronomical detections of the 
three doubly $^{13}$C-substituted isotopologues of ethyl cyanide toward 
Sgr~B2(N2).

\end{appendix}


\end{document}